\documentclass[a4paper,11pt]{article}

\usepackage[a4paper,top=3.5cm,right=3.1cm,left=3.1cm,bottom=3.5cm]{geometry}
\usepackage{amsthm}
\usepackage{bbm}

\usepackage{hyperref}
\usepackage{latexsym,amsfonts,amsmath,amssymb,enumerate,epsfig}
\usepackage{mathdots}
\usepackage{subfig}
\usepackage{mathrsfs}
\usepackage{ulem}

\usepackage{color}

\numberwithin{equation}{section}
\newtheorem{definition}{Definition}[section]

\newtheorem{proposition}[definition]{Proposition}

\newtheorem{lemma}[definition]{Lemma}
\newcommand{\mb}[1]{\quad\mbox{#1}\quad}
\newcommand{\beq}{\begin{equation}}
\newcommand{\eeq}{\end{equation}}
\newcommand{\bea}{\begin{eqnarray}}
\newcommand{\eea}{\end{eqnarray}}
\newcommand{\beano}{\begin{eqnarray*}}
\newcommand{\eeano}{\end{eqnarray*}}

\newcommand{\finproof}{{\hfill \rule{5pt}{5pt}\\}}

        \def\cR{{\cal R}}
\def\cS{{\cal S}}

\def\fm{{\mathfrak m}}

\def\ft{{\mathfrak t}}

\newcommand{\CC}{{\mathbb C}}

\newcommand{\wb}[1]{{\overline{#1}}}

\newcommand{\wt}[1]{{\widetilde{#1}}}

\newcommand{\ato}[2]{\genfrac{}{}{0pt}{}{#1}{#2}}

\begin{document}
\setcounter{page}{0}
\pagestyle{empty}
\null

\vfill

\begin{center}
{\LARGE  {\sffamily Integrable boundary conditions for multi-species ASEP
}}
\vspace{60pt}

\begin{large}
 N. Crampe$^{a}$\footnote{nicolas.crampe@umontpellier.fr},
 C. Finn$^{b}$\footnote{caley.finn@lapth.cnrs.fr},
 E. Ragoucy$^{b}$\footnote{eric.ragoucy@lapth.cnrs.fr}
 and M. Vanicat$^{b}$\footnote{matthieu.vanicat@lapth.cnrs.fr}
 \\[.41cm] 
$^a$ Laboratoire Charles Coulomb (L2C), CNRS - Universit{\'e} de Montpellier,\\
Place Eug\`ene Bataillon, CC069,
F-34095 Montpellier Cedex 5, France.
\\[.42cm]
$^{b}$  LAPTh,
 CNRS - Universit{\'e} Savoie Mont Blanc\\
   9 chemin de Bellevue, BP 110, F-74941  Annecy-le-Vieux Cedex, 
France. 
\end{large}
\end{center}

\vfill

\begin{abstract}
The first result of the present paper is to provide classes of 
explicit solutions for integrable boundary matrices for the multi-species ASEP with an arbitrary number of species.

All the solutions we have obtained can be seen as representations of a new algebra that contains the boundary Hecke algebra. 
The boundary Hecke algebra is not sufficient to build these solutions.
This is the second result of our paper.

\end{abstract}
\vfill
\vfill
\begin{flushright}
June 2016\\
LAPTH-025/16
\end{flushright}
\vfill
\newpage
\pagestyle{plain}

\setcounter{footnote}{0}

\section{Introduction}

The asymmetric simple exclusion process (ASEP)\cite{Spitzer,Liggett} describes particles that hop on a one-dimensional lattice with anisotropic rates
and hard core exclusion. Though it is one of the simplest examples of a driven diffusive system, it has become over the last decades 
a paradigm in out-of-equilibrium statistical physics \cite{Derrida98}. It displays indeed a rich phenomenology (such as boundary induced phase transitions, 
shock waves,...)
and has found many applications in biology and traffic flow \cite{CMZ,CSS}. A particularly remarkable feature of this stochastic process is that it is
integrable. It has thus attracted much interest in combinatorics, mathematical physics and probability theory.

The bulk dynamics of the ASEP can be generalised to several species of particles, preserving the integrability property. It has led to   
many studies, among them can be mentioned the computation of the stationary state, for the model with periodic boundary conditions, using
a matrix product ansatz \cite{pem,CGW,KMO}. Then, the stationary state has been computed for reflective boundaries \cite{arita} and 
semi-permeable boundaries \cite{Uchiyama}.
The case of the multi-species system with generic boundaries, which is of particular 
interest in out-of-equilibrium statistical physics for the comprehension of boundary induced phenomenon, appears more complicated.    

Fortunately, the integrability property allows one to 
choose the particles injection and extraction rates at the boundaries which permits the computation of the stationary state \cite{SW,CRV}.
Indeed, the inverse scattering method provides a general framework to
determine the boundary conditions that preserve the integrability of the model \cite{skl}. 
However, the price to pay consists in solving a compatibility equation between the dynamics of the bulk and
the reflection rates, called the reflection equation.
The resolution of this equation is a complicated problem and is at the heart of a lot of research (see e.g. \cite{vega,abad,BFKZ,gan,AACDFR}).

Recent progress has been made to classify the integrable boundaries for the two-species ASEP and to compute the associated stationary 
state in a matrix product form \cite{CMRV}. The goal of this paper is to provide integrable boundaries for the multi-species ASEP with an arbitrary number of species.
The integrable boundaries that we find divide the set of all species into five classes, which we call very-slow, slow, intermediate, fast and very-fast species. 
This division is labeled by four integers. There are also two free real parameters, associated to transition rates on the boundaries.
We show also that all these integrable boundaries satisfy a generalization of the boundary Hecke algebra. 
This generalisation is necessary to take into account the whole set of the solutions we found and, to our knowledge, is new in the literature.

The outline of this paper is as follows. In section 2 we recall briefly the stochastic dynamics of the multi-species ASEP and the quantum
inverse scattering framework used to determine the integrable boundary conditions. Section 3 is devoted to the presentation of a class of 
integrable boundaries on the left and on the right. We also point out examples of combination of boundaries for which the Markov chain is
irreducible. In section 4 we introduce a novel algebra, which includes all the boundary matrices presented.
This algebra then allows construction of $K$-matrices solving the reflection equation 
 through an easy Baxterisation procedure. 
We argue that some of the boundary conditions presented in section 3 do not fit in the standard framework of boundary Hecke algebras and that 
the algebraic structure defined in this paper is needed to encompass them.

\section{Multi-species ASEP}

\subsection{Presentation of the model}
  We start  by  recalling  the dynamical rules of the multi-species ASEP.  We will call the model multi-species ASEP, with the convention 
  that we consider a model with $(N-1)$ species of particles on a one-dimensional lattice with $L$ sites. Each  site on the lattice 
  is occupied by a single particle, or is empty, and we identify this vacancy (or hole) as an additional species, that we call 1. 
  The species of particles shall be labeled $2,3,\dots,N$. 
  A configuration on the lattice is thus a $L$-tuplet $\big(\tau_1,\tau_2,...,\tau_L\big)$ that belongs to 
$\{1,\ldots,N\}^L$. To each of the $L$ sites we will associate a $\CC^N$ vector space, so that the set of all configurations is embedded into the tensor space 
$\underbrace{\CC^N\otimes\cdots\otimes\CC^N}_{L}$. The natural basis of this space is given by 
$|\tau_1\rangle\otimes \cdots \otimes |\tau_L\rangle$ with $\tau_i=1,2,...,N$ and $|\tau\rangle=(\underbrace{0,\dots,0}_{\tau-1},1,\underbrace{0,\dots,0}_{N-\tau})^t$.

\paragraph{In the bulk.}  
  The dynamics is defined as follows: A  bond $(i,i+1)$,  with $
  1 \le i \le L-1$, between two neighboring lattice sites, is updated
  between time $t$ and $t +dt$ by swapping the particles  at $i$ and
  $i+1$ with rate $1$ or $q$ depending on the local configuration $ \tau_i \, \tau_{i+1}$ involved
\beq
\begin{aligned}
&  \tau_i \, \tau_{i+1}\, \xrightarrow{\ 1\ } \,\tau_{i+1}\,\tau_i \mb{if}  \tau_i > \tau_{i+1}\,,  \\
&  \tau_i \, \tau_{i+1}\, \xrightarrow{\ q\ } \,\tau_{i+1}\,\tau_i \mb{if}\tau_i < \tau_{i+1}\,,
\end{aligned}
\eeq
 where $q$ is a free positive parameter. 
   These rules show that particles are ordered by their species: the species $N$  has the highest
   priority, followed by species $(N-1)$, down to particles of species $2$, and lastly by holes (i.e. species 1). 
Species with higher priority will be said to be faster, so that species $N$ is the fastest species (it is the
flash) and species 1 the slowest.\footnote{Note that this interpretation makes sense when $q < 1$.}

The bulk rules can be encoded in a local Markov matrix acting on two sites, i.e. on $\CC^N\otimes\CC^N$. Explicitly, it has the form
\beq\label{eq:mlocal}
\fm=\sum_{1\leq i<j\leq N} \left\{\Big(E_{ij}\otimes E_{ji}-E_{jj}\otimes E_{ii}\Big)+
q\Big(E_{ji}\otimes E_{ij}-E_{ii}\otimes E_{jj}\Big)\right\},
\eeq
where $E_{ij}$ is the $N\times N$ elementary matrix with 1 at position $(i,j)$ and 0 elsewhere.
This matrix can be obtained from an $R$-matrix satisfying the Yang-Baxter equation which allows us to prove integrability of the model: 
this construction will be briefly recalled in the next section.
The complete Markov matrix in the bulk is given by
\beq\label{Mbulk}
M_{bulk}=\fm_{12}+\fm_{23}+\dots+\fm_{L-1,L}\,,
\eeq
where the indices on $\fm$ indicate on which copies of $\CC^N$ it acts non-trivially.

\paragraph{On the boundaries.} Particles are allowed to enter or to exit from both boundaries and
 the corresponding  entrance/exit rates  may depend on the type of the
 particle that was previously located at the boundary. More precisely,
 both on the left and on the right boundary, we can have  a transition
 of the type 
 \beq
 \tau_1 \, \xrightarrow{\, r(\tau_1,\tau_2)\, } \,\tau_2\;,
 \eeq
 for $\tau_1,\tau_2=1,2,\dots,N$.
This
 leads to $2N(N-1)$ independent rates (that is ${N(N-1)}$ rates on each side). The rates corresponding to the left boundary are gathered in an $N\times N$ boundary matrix $B$:
 \beq
 B=\sum_{1\leq i\neq j\leq N} r(\tau_i,\tau_j)\,E_{ji} - \sum_{i=1}^N \Big(\sum_{j\neq i}r(\tau_i,\tau_j)\Big) E_{ii}
 .
 \eeq
 Similarly, the rates for the right boundary are gathered in a matrix  $\wb B$. 
 An open multi-species ASEP will be defined by the bulk matrix \eqref{Mbulk} and the two boundary matrices $B$ and $\wb B$. 
The Markov matrix associated to the model will be 
\beq
M=M_{bulk} + B_1+ \wb B_L,
\eeq
and the master equation, governing the time evolution of the probability $P_t(\tau_1,\dots,\tau_L)$ to be in
the configuration $(\tau_1,\dots,\tau_L)$, is written
\beq
\frac{d  |P_t\rangle}{ dt}  =M |P_t\rangle\mb{where}
 |P_t\rangle = \sum_{1\leq\tau_1,\dots,\tau_L\leq N} P_t(\tau_1,\dots,\tau_L)\,|\tau_1\rangle\otimes \cdots \otimes |\tau_L\rangle\;.
\eeq

Although the bulk part $M_{bulk}$ corresponds to an integrable model, for arbitrary choices of the boundary rates, the model will not be integrable. 
The first result of the present paper is to provide classes of 
explicit solutions for integrable boundary matrices for the multi-species ASEP. They are presented in section \ref{sect:solu}. 

\subsection{Integrable approach to open models}

We briefly recall the context of the quantum inverse scattering method that allows one to define open integrable models. 
We refer to the historical paper \cite{skl} and to the review \cite{CRV} for more details.

We introduce an $R$-matrix acting on two copies of $\CC^N$. It obeys the  Yang-Baxter equation 
and the unitarity relation:
\bea
&& {R}_{12}\left(\frac{x_1}{x_2}\right)\, {R}_{13}\left(\frac{x_1}{x_3}\right)\, {R}_{23}\left(\frac{x_2}{x_3}\right)=
    {R}_{23}\left(\frac{x_2}{x_3}\right)\, {R}_{13}\left(\frac{x_1}{x_3}\right)\, {R}_{12}\left(\frac{x_1}{x_2}\right)\;,
\label{ybe}
\\
&&{R}_{12}(x)\, {R}_{21}\left(\frac1x\right) = 1\;.\label{Runit}
\eea
Again, the indices indicate in which copies of $\CC^N$ the $R$-matrices act non-trivially.
For the multi-species ASEP, the R-matrix can be written as follows
\begin{equation}\label{eq:Rasep}
 R_{i,i+1}(x)=P_{i,i+1}(1+ \frac{x-1}{q x - 1}\, \fm_{i,i+1})
\end{equation}
where $P$ is the permutation operator that exchanges the two copies of $\CC^N$ in $\CC^N\otimes \CC^N$.

To define an integrable open model, one introduces the transfer matrix \cite{skl}:
\beq\label{eq:transf-open}
\ft_{open}(x)=tr_0\Big({R}_{0L}(x)...{R}_{01}(x)\,K_0(x)\,{R}_{10}(x)...{R}_{L0}(x)\wt K_0(x)\Big)\,,
\eeq
where $K(x)$ is a $N\times N$ matrix which obeys the reflection equation
and is unitary:
\bea
&& {R}_{12}\left(\frac{x_1}{x_2}\right) K_1(x_1) {R}_{21}(x_1 x_2) K_2(x_2)
    =  K_2(x_2) {R}_{12}(x_1 x_2) K_1(x_1) {R}_{21}\left(\frac{x_1}{x_2}\right)\,,\quad
\label{eq:reflx}
\\
&&  K(x) K\left(\frac1x\right) = 1.\label{eq:unitK}
\eea
The boundary matrix $\wt K(x)$ in \eqref{eq:transf-open} satisfies a dual reflection equation. The solutions
to this dual reflection equation
can be obtained from the solutions $K(x)$ of the reflection equation \eqref{eq:reflx}  by 
\begin{equation}
 \wt{K_1}(x) =
tr_0\left(\wb K_0(1/x)\,  ((R_{01}(x^2)^{t_1})^{-1})^{t_1} \, P_{01}\right)
\end{equation}
where 
\beq\label{def:Kbar}
\wb K(x)=U K\left(1/x\right)\,U \mb{and} U=\begin{pmatrix}
          &&1\\
          &\iddots&\\
          1&&
         \end{pmatrix}.
\eeq

From these properties, usual calculations \cite{skl} prove that the transfer matrix 
$\ft_{open}(x)$ defines an integrable model: $[\ft_{open}(x)\,,\,\ft_{open}(y)]=0$. The global Markov matrix is then defined as
\beq
M=\frac{q-1}{2} \left.\frac{d}{dx}\ft_{open}(x)\right|_{x=1}\;.
\eeq
Then the integrable boundaries are obtained from the K-matrices by 
\beq\label{eq:KB}
B=\frac{q-1}{2} \left.\frac{d}{dx}K(x)\right|_{x=1}\mb{and}
\wb B=-\frac{q-1}{2} \left.\frac{d}{dx}\wb K(x)\right|_{x=1}\;.
\eeq

\section{Integrable boundary conditions for the multi-species ASEP}

\subsection{Presentation of the left boundary conditions/matrices\label{sect:solu}}

We wish to give explicit solutions for integrable Markovian boundary matrices for the multi-species ASEP. 
These solutions are obtained with relations \eqref{eq:KB} from $K$-matrices obeying the reflection equation \eqref{eq:reflx} with the R-matrix \eqref{eq:Rasep}.
We present here the integrable Markovian boundary conditions. We postpone the presentation of the associated
K-matrices to section \ref{sect:Bdecomp}
and proof of the integrability to section \ref{sec:nasepInteg}.

The integrable boundary conditions depend on two free real positive parameters (rates) $\alpha$ and $\gamma$, and four positive integers 
$s_1,\ s_2,\ f_1$ and $f_2$, that label two special slow ($s$) and two special fast ($f$) species, with the conditions
\beq
1\leq s_1\leq s_2<f_2\leq f_1\leq N\mb{and}f_1-f_2 = s_2-s_1.\label{cont:sf}
\eeq 
The four special species will be essentially created on the boundary, while the remaining species will
essentially (but not only) decay onto these four types. Any species in between $s_1$ and $s_2$ will be paired
with one species in between $f_2$ and $f_1$,  allowing a transmutation (on the boundary) between the pairs.
Finally, in between $s_2$ and $f_2$, either nothing happens, or the species decay to $s_2$ and $f_2$.
\pagebreak[3]

More specifically, integrability is preserved when we  have the following rules and rates on the boundary:
\begin{itemize}
\item \textbf{Class of very slow species:} for species $\tau$ with $1\leq \tau<s_1$, we have:
\beq
\tau\ \raisebox{1ex}{$\ato{\gamma}{\longrightarrow}$}\ s_1 \mb{and} \tau\ \raisebox{1ex}{$\ato{\alpha}{\longrightarrow}$}\ f_1.
\eeq
\item  \textbf{Class of slow species:} for species $\tau$ with $s_1\leq \tau\leq s_2$, we have:
\beq
\tau\ \raisebox{1ex}{$\ato{\alpha}{\longrightarrow}$}\ \wb \tau=s_1+f_1-\tau=s_2+f_2-\tau.
\eeq
\item  \textbf{Class of intermediate species:} for species $\tau$ with $s_2< \tau<f_2$, we have the two possibilities:
\begin{enumerate}
\item $\tau\ \raisebox{1ex}{$\ato{0}{\longrightarrow}$}\ \tau'$, $\forall \tau'$ (no decay, creation or transmutation).
\item $\tau\ \raisebox{1ex}{$\ato{\wt\gamma}{\longrightarrow}$}\ s_2$ and $\tau\ \raisebox{1ex}{$\ato{\alpha}{\longrightarrow}$}\ f_2$.
\end{enumerate}
\item \textbf{Class of  fast species:} for species $\tau$ with $f_2\leq \tau\leq f_1$, we have:
\beq
\tau\ \raisebox{1ex}{$\ato{\wt\gamma}{\longrightarrow}$}\ \wb \tau=s_1+f_1-\tau.
\eeq
\item \textbf{Class of very fast species:} for species $\tau$ with $f_1< \tau\leq N$, we have:
\beq
\tau\ \raisebox{1ex}{$\ato{\wt\gamma}{\longrightarrow}$}\ s_1 \mb{and} \tau\ \raisebox{1ex}{$\ato{\wt\alpha}{\longrightarrow}$}\ f_1.
\eeq
\end{itemize}
We have introduced the following combination of the rates:
\begin{equation}\label{eq:tilde}
    \tilde{\alpha} = \frac{(\alpha + \gamma + q - 1)\alpha}{\alpha + \gamma},
    \qquad
    \tilde{\gamma} = \frac{(\alpha + \gamma + q - 1)\gamma }{\alpha + \gamma}.
\end{equation}
This implies that $\alpha$, $\gamma$, $q$ are constrained such that $\tilde\alpha$, $\tilde\gamma$ are
positive.

Note that, depending on the choice of $s_1$, $s_2$, $f_2$ and $f_1$, some classes of species may not occur: for instance if $s_1=1$, there is no very 
slow species. In the same way, if $f_2=s_2+1$, there are no intermediate species.

Due to the second constraint in \eqref{cont:sf}, the number of slow species coincides with the number of fast species, in accordance with the pairing mentioned above.
By counting the number of possibilities for $s_1$, $s_2$, $f_1$ and $f_2$ with the constraints \eqref{cont:sf}, we can deduce that, for
multi-species ASEP there exist\footnote{We have included in the counting the two possible choices for the intermediate species when $f_2>s_2+1$.} 
$\begin{pmatrix}  N+1\\  3   \end{pmatrix}$ 
different integrable boundaries, each of them depending on two real parameters.

Note that in any transition, the number of particles for the species in the very slow and very fast classes can only decrease. 
It may stay constant for the slow, fast and intermediate classes. For the four special types
it may increase.\\

To summarize, these rates are gathered in the two following type of boundary matrices, depending on the two possibilities for intermediate species:
\bea
&&B^0(\alpha,\gamma|s_1, s_2, f_2, f_1)=
\label{eq:Bempty}\\
&&\left(
\begin{array}{ccc|cccc|ccc|cccc|ccc}
\mbox{-}\sigma & & & & & & & & & & & & & & & &
\\
 &\ddots & & & & & & & & & & & & & & &
\\
 & & \mbox{-}\sigma & & & & & & & & & & & & & &
\\
\hline
\gamma & \cdots&\gamma & \mbox{-}\alpha & & & & & & & & & &\wt\gamma & \wt\gamma& \cdots & \wt\gamma
\\
 & & & & \mbox{-}\alpha& & & & & & & &\wt\gamma & & & & 
\\
 & & & & &\ddots & & & & & &\iddots & & & & & 
\\
 & & & & & & \mbox{-}\alpha & & & & \wt\gamma& & & & & & 
\\
\hline
 & & & & & & &0 &\cdots& 0 & & & & & & & 
\\
 & & & & & & &\vdots & & \vdots & & & & & & & 
\\
 & & & & & & &0 &\cdots& 0 & & & & & & & 
\\
\hline
 & & & & & & \alpha& & & & \mbox{-}\wt\gamma& & & & & & 
\\
 & & & & &\iddots & & & & & &\ddots & & & & & 
\\
 & & & &\alpha & & & & & & & & \mbox{-}\wt\gamma& & & & 
\\
\alpha & \cdots& \alpha& \alpha& & & & & & & & & &\mbox{-}\wt\gamma & \wt\alpha& \cdots& \wt\alpha
\\
\hline
 & & & & & & & & & & & & & & \mbox{-}\wt\sigma & &
\\
 & & & & & & & & & & & & & & & \ddots&
\\
 & & & & & & & & & & & & & & & &  \mbox{-}\wt\sigma
\end{array}
\right)\nonumber
\eea

\bea
&&B(\alpha,\gamma|s_1, s_2, f_2, f_1)=
\label{eq:Bfull}\\
&&\left(
\begin{array}{ccc|cccc|ccc|cccc|ccc}
\mbox{-}\sigma & & & & & & & & & & & & & & & &
\\
 &\ddots & & & & & & & & & & & & & & &
\\
 & & \mbox{-}\sigma & & & & & & & & & & & & & &
\\
\hline
\gamma & \cdots&\gamma & \mbox{-}\alpha & & & & & & & & & &\wt\gamma & \wt\gamma& \cdots & \wt\gamma
\\
 & & & & \mbox{-}\alpha& & & & & & & &\wt\gamma & & & & 
\\
 & & & & &\ddots & & & & & &\iddots & & & & & 
\\
 & & & & & & \mbox{-}\alpha & \wt\gamma & \cdots & \wt\gamma & \wt\gamma& & & & & & 
\\
\hline
 & & & & & & & \mbox{-}\sigma' & & & & & & & & & 
\\
 & & & & & & & & \ddots & & & & & & & & 
\\
 & & & & & & & & & \mbox{-}\sigma' & & & & & & & 
\\
\hline
 & & & & & & \alpha & \alpha & \cdots & \alpha & \mbox{-}\wt\gamma& & & & & & 
\\
 & & & & &\iddots & & & & & &\ddots & & & & & 
\\
 & & & &\alpha & & & & & & & & \mbox{-}\wt\gamma& & & & 
\\
\alpha & \cdots& \alpha& \alpha& & & & & & & & & &\mbox{-}\wt\gamma & \wt\alpha& \cdots& \wt\alpha
\\
\hline
 & & & & & & & & & & & & & & \mbox{-}\wt\sigma & &
\\
 & & & & & & & & & & & & & & & \ddots&
\\
 & & & & & & & & & & & & & & & &  \mbox{-}\wt\sigma
\end{array}
\right)\nonumber
\eea
We have introduced $\sigma=\alpha+\gamma$, $\wt\sigma=\wt\alpha+\wt\gamma$ and $\sigma'=\alpha+\wt\gamma$.
The empty spaces in the matrices above are filled with zeros, and the lines indicate the positions of the four special types of species.

\paragraph{More solutions:} One can produce more integrable solutions using conjugation by any diagonal invertible matrix $V$. Indeed, due to the invariance 
of the R-matrix \eqref{eq:Rasep} by the conjugation by $V_1V_2$, $VK(x)V^{-1}$ is solution of the reflection equation if $K(x)$ is also a solution. However,
the resulting conjugated matrix may not be Markovian. Nonetheless, we remark that conjugation by the diagonal matrix 
$diag(e^{s_1},e^{s_2},\dots,e^{s_N})$ provides a deformed integrable boundary matrix that 
allows one to compute the cumulants of the currents at the boundary for the different species. 

\subsection{Construction of the right boundary matrices}

A right boundary matrix $\wb B$ is obtained through the relation \eqref{eq:KB}, where $\wb K(x)$ is deduced from a solution $K(x)$ thanks to \eqref{def:Kbar}. Let us stress that the parameters entering $\wb B$ are independent from the ones used in the left boundary $B$. Altogether we will have four real parameters: $\alpha,\gamma$ for the left boundary, and $\beta,\delta$ for the right one. In the same way, the labels $s_1', s_2', f_2', f_1'$ of the four special species in the right boundary are independent from the four special species labels $s_1, s_2, f_2, f_1$ in the left boundary. Explicitly, the right boundary matrices are defined as
\beq
\wb B(\beta,\delta|s_1', s_2', f_2', f_1') = U\,B(\beta,\delta|s''_1, s''_2, f''_2, f''_1)\,U^{-1}
\eeq
where $U$ is defined in \eqref{def:Kbar}. The conjugation by $U$ implies
$f''_j=N+1-s'_j$ and $s''_j=N+1-f'_j$, $j=1,2$.

The bijection between right and left boundaries can be seen in the following identity
\beq
\wb B(\beta,\delta|s_1, s_2, f_2, f_1)\equiv  B(\beta,\delta|s_1, s_2, f_2, f_1)\Big|_{z\leftrightarrow \wt z}
\eeq
where $z\leftrightarrow \wt z$  means that we interchange $\beta$ with $\wt \beta$ and $\delta$ with $\wt\delta$. 
As in the case of left boundaries, we use the notation
\beq
\wt \beta =  \frac{(\beta + \delta + q - 1)\beta }{\beta + \delta}
\mb{and}\wt \delta =  \frac{(\beta + \delta + q - 1)\delta }{\beta + \delta}.
\eeq

\subsection{Examples} 
\paragraph{For the case $N=2$, we recover the one-species ASEP.} We get only one possible choice for $s_1$, $s_2$, $f_1$ and $f_2$ 
given by $s_1=s_2=1$ and $f_1=f_2=2$. Then, in the language used in this paper, the particle 1 (vacancy) is slow and the particle 2 is fast
and the rates at the boundary are given by 
\beq
1\ \raisebox{1ex}{$\ato{\alpha}{\longrightarrow}$}\ 2
\mb{and} 2\ \raisebox{1ex}{$\ato{\wt\gamma}{\longrightarrow}$}\ 1.
\eeq
 One recovers that for the one-species ASEP,
the generic boundary is integrable. The boundary matrix has the form
\beq
B=\begin{pmatrix} -\alpha & \wt\gamma \\ \alpha & -\wt\gamma \end{pmatrix}.
\eeq
One can use Bethe ansatz method to compute the eigenvalues and compute for example the spectral gap \cite{gier}.

Conjugation by a diagonal matrix provides the non-Markovian boundary matrix used to compute the cumulant of the 
current \cite{cumulants}:
\beq
B(s)=\begin{pmatrix} -\alpha & e^s\,\wt\gamma \\ e^{-s}\,\alpha & -\wt\gamma \end{pmatrix}.
\eeq
It still corresponds to an integrable boundary.
 
\paragraph{For the case $N=3$, we obtain the two-species ASEP.} There are four possibilities summarized in table \ref{table:N3}. We recover the boundaries found in \cite{CMRV}.
\begin{table}[htb] 
\begin{center}
   \begin{tabular}{|c|c|c|c|c|}
     \hline
     &$s_1=s_2=1$ & $s_1=s_2=2$  &  \multicolumn{2}{c|}{$s_1=s_2=1$}     \\
     &$f_1=f_2=2$ & $f_1=f_2=3$  &  \multicolumn{2}{c|}{$s_1=s_2=3$ }    \\
     \hline
\begin{tabular}{c} Type of  \\   part. \end{tabular}  
&\begin{tabular}{c} part. 1 slow  \\ part. 2 fast \\ part. 3 very fast \end{tabular}  
& \begin{tabular}{c} part. 1 very slow \\ part. 2 slow \\ part. 3 fast \end{tabular}  
& \multicolumn{2}{c|}{\begin{tabular}{c}part. 1 slow \\ part. 2 intermediate \\ part. 3 fast \end{tabular}  }\\
     \hline
    & \rule{0ex}{3.5ex}1\ \raisebox{1ex}{$\ato{\alpha}{\longrightarrow}$}\ 2    & 1\ \raisebox{1ex}{$\ato{\gamma}{\longrightarrow}$}\ 2
      && 1\ \raisebox{1ex}{$\ato{\alpha}{\longrightarrow}$}\ 3 \\
Rates     & 2\ \raisebox{1ex}{$\ato{\wt\gamma}{\longrightarrow}$}\ 1& 1\ \raisebox{1ex}{$\ato{\alpha}{\longrightarrow}$}\ 3
      &  1\ \raisebox{1ex}{$\ato{\alpha}{\longrightarrow}$}\ 3& 2\ \raisebox{1ex}{$\ato{\wt\gamma}{\longrightarrow}$}\ 1\\
     & 3\ \raisebox{1ex}{$\ato{\wt\gamma}{\longrightarrow}$}\ 1& 2\ \raisebox{1ex}{$\ato{\alpha}{\longrightarrow}$}\ 3
      & 3\ \raisebox{1ex}{$\ato{\wt\gamma}{\longrightarrow}$}\ 1& 2\ \raisebox{1ex}{$\ato{\alpha}{\longrightarrow}$}\ 3\\
     & 3\ \raisebox{1ex}{$\ato{\wt\alpha}{\longrightarrow}$}\ 2& 3\ \raisebox{1ex}{$\ato{\wt\gamma}{\longrightarrow}$}\ 2
      &&3\ \raisebox{1ex}{$\ato{\wt\gamma}{\longrightarrow}$}\ 1\\
      \hline
     \rule{0ex}{3ex}Name in \cite{CMRV}&$L_1$&$L_2$&$L_4$&$L_3$\\
      \hline
   \end{tabular}
   \caption{\label{table:N3} The four integrable boundaries in the case N=3. The last row corresponds 
   to the names of these boundaries in \cite{CMRV}.}
\end{center}
\end{table}

\paragraph{Generic examples.} Some of the boundary matrices can be related to former studies 
of boundary Hecke algebras (see also section \ref{sec:heckeb}). In our notation, they correspond to 
the matrices $B(\alpha,\gamma|1,s_2,N+1-s_2,N)$ or $B^0(\alpha,\gamma|1,s_2,N+1-s_2,N)$. Among them, some have been 
considered: $B^0(\alpha,\gamma|1,2,N-1,N)$ was analyzed in \cite{PhD-Corteel}, 
and for the two-species ASEP ($N=3$) $B^0(\alpha,\gamma|1,1,3,3)$ was studied in \cite{Uchiyama,Corteel,Cantini}.

\subsection{Irreducible open multi-species ASEP\label{sect:irred}}

Since the boundary matrices we exhibited depend only on two different rates, one can wonder if, when using these boundaries, the open multi-species ASEP ``trivialises'' for $N$ big enough. More precisely, one may ask whether
 there is some limit on the number of species above which a multi-species ASEP can always be mapped (through identification)  to an ASEP with a smaller number of species. 
 In fact, it is not the case, thanks to the four types of special species, that can be chosen freely on each of the two boundaries. 
 Indeed it can be shown that there are pairs of boundary matrices for which the number of particles of any given species  is not conserved. 
 Moreover, for any given subset of species, the total number of particles whose species is in this subset is not conserved either. 

We give below examples of such pairings of boundary matrices. We write them 
as $B=B(\alpha,\gamma|s_1, s_2, f_2, f_1)$ and $\wb B=\wb B(\beta,\delta|s'_1, s'_2, f'_2, f'_1)$, 
where the first matrix represents the left boundary, and the second matrix the right one. 
The explicit values of the four special species (for each boundary) 
depends on the parity of $N$ in the multi-species ASEP:
\paragraph{For the multi-species ASEP with $N=2n+1$,} we can consider the matrices $B=B(\alpha,\gamma|2,n+1,n+2,2n+1)$ and 
$\wb B=\wb B(\beta,\delta|1, n, n+1, 2n)$. Explicitly, they are given by
\bea
B &=&
\left(
\begin{array}{c|cccccc}
\mbox{-}\sigma & & & & & & 
\\
\hline
\gamma & \mbox{-}\alpha & & & & &\wt\gamma 
\\
 & & \ddots & & &\iddots & 
\\
 & &  & \mbox{-}\alpha & \wt\gamma& & 
\\
 & & &  \alpha & \mbox{-}\wt\gamma & 
\\
 & & \iddots & & &\ddots &
\\
\alpha & \alpha&  & & & &\mbox{-}\wt\gamma
\end{array}
\right)\mb{with} \sigma=\alpha+\gamma
\nonumber
\eea

\bea
\wb B &=&
\left(
\begin{array}{cccccc|c}
 \mbox{-}\wt\delta & & & & &\beta & \beta
\\
 & \ddots & & &\iddots & &  
\\
 &  & \mbox{-}\wt\delta & \beta& &  &  
\\
 & &  \wt\delta & \mbox{-}\beta& &  &  
\\
  &\iddots & & &\ddots & &  
\\
 \wt\delta&  & & & &\mbox{-}\beta & \delta
\\
\hline
 & & & & & & \mbox{-}\sigma 
\end{array}
\right) \mb{with} \sigma=\beta+\delta
\nonumber
\eea

In both cases, the intermediate particles drop out because we choose $f_2=s_2+1$ and the very fast (resp. very slow) particles do not exist
in $B$ (resp. in $\wb B$). Then, we have drawn only the line corresponding to $s_1$ in $B$ and to $f_1$ in $\wb B$.

The evolution of the system given by the Markov chain with these boundaries does not preserve the number of particles of any subset of species.
To prove that, we can see that there exists a cycle that connects all the species of the particles and the holes :
\beq
\begin{array}{cccccccccccc}
1 & \xrightarrow{\ \alpha\ } & 2n+1 & \xrightarrow{\ \wt\gamma\ } & 2 & \xrightarrow{\ \wt\delta\ } & 2n-1 & \xrightarrow{\ \wt\gamma\ } & 4
& \xrightarrow{\ \wt\delta\ } & \dots &
\\
\ \uparrow \scriptstyle{\beta}&&&&&&&&&&&\vdots\\
2n & \xleftarrow{\ \alpha\ } & 3 & \xleftarrow{\ \beta\ } & 2n-2 & \xleftarrow{\ \alpha\ } & 5 & \xleftarrow{\ \beta\ } & 2n-4
& \xleftarrow{\ \alpha\ } & \dots&
\end{array}
\eeq

\paragraph{For the multi-species ASEP with $N=2n+2$,} we can consider the matrices $B=B(\alpha,\gamma|2,n+1,n+3,2n+2)$ and
$\wb B=\wb B(\beta,\delta|1,n,n+2,2n+1)$, namely
\bea
&&B=
\left(
\begin{array}{c|ccc|c|ccc}
 \mbox{-}\sigma & & & & & & &  
\\
\hline
\gamma & \mbox{-}\alpha & & & & & &\wt\gamma 
\\
 & & \ddots & & & &\iddots & 
\\
 & & &\mbox{-}\alpha & \wt\gamma & \wt\gamma & & 
\\
\hline
 & &  & & \mbox{-}\sigma' & & &
\\
\hline
 & & &\alpha & \alpha & \mbox{-}\wt\gamma & & 
\\
 & & \iddots & & & &\ddots & 
\\
 \alpha& \alpha& & & & & &\mbox{-}\wt\gamma 
\end{array}
\right)\mb{with}
\begin{cases} \sigma=\alpha+\gamma\\ \sigma'=\alpha+\wt\gamma \\ \wt\sigma=\wt\alpha+\wt\gamma
\end{cases}
\nonumber
\eea

\bea
&&\wb B=
\left(
\begin{array}{ccc|c|ccc|c}
 \mbox{-}\wt\delta & & & & & &\beta & \beta 
\\
 & \ddots & & & &\iddots & &  
\\
 & &\mbox{-}\wt\delta & \beta & \beta & & & 
\\
\hline
 &  & & \mbox{-}\sigma' & & & &  
\\
\hline
\rule{0ex}{3ex} & &\wt\delta & \wt\delta & \mbox{-}\beta & & & 
\\
 & \iddots & & & & \ddots & &  
\\
 \wt\delta& & & & & & \mbox{-}\beta & \delta
\\
\hline
 & & & & & & & \mbox{-}\sigma 
\end{array}
\right)\mb{with}
\begin{cases} \sigma=\beta+\delta \\ \sigma'=\beta+\wt\delta 
\end{cases}
\nonumber
\eea
One sees that now the very fast species have been dropped from $B$ and the very slow species from $\wb B$.

Again, there exists a cycle (similar to the one above) that connects all the species and the holes, however its form for the intermediate 
species depends on the parity of $n$.

\subsection{$K$-matrix\label{sect:Bdecomp}}

To make contact with $K$-matrices and integrability, we decompose both matrices \eqref{eq:Bempty} or \eqref{eq:Bfull} into three pieces, with
\beq
b_0^+=\left(\begin{array}{ccc|c|c|c|c} \mbox{-}\gamma& & & & & & \\ & \ddots& & & & & \\ & &\mbox{-}\gamma & & & & \\ 
\hline \gamma& \cdots& \gamma& & & & \\ 
& & & & & & \\ \hline & & & & & & \\ \hline & & & & & &\\ \hline & & & & & & \end{array}\right)
\mb{and}
b_0^-=\left(\begin{array}{c|c|c|c|ccc} & & & & & & \\ \hline & & & & & & \\ \hline & & & & & & \\ 
\hline& & & & & & \\ 
& & & & \wt\alpha& \cdots&\wt\alpha \\  \hline & & & & \mbox{-}\wt\alpha& & \\ & & & & & \ddots & \\  & & & & & & \mbox{-}\wt\alpha
\end{array}\right)
\label{eq:bpm}
\eeq
where we draw symbolically the lines corresponding to the four special types of particles, to indicate which part of the matrix we picked up in the boundary matrix to construct $b_0^\pm$. 
Again, the empty spaces are all filled with zeros. The remaining part is 
$b_0=B-(b_0^+ + b_0^-)$, where $B$ is either \eqref{eq:Bempty} or \eqref{eq:Bfull}.
Note that the decomposition is done in such a way that each matrix $b_0$, $b_0^\pm$ is Markovian.

This decomposition of the boundary $B=b_0+b_0^++b_0^-$ allows the associated K-matrix to be written as
\begin{align}
    & K(x) = 1 + k(x)\,\Big(b_0 + x \,b_0^+ + \frac1x \,b_0^-\Big),\label{K:genHecke}\\
    & \mbox{with}\quad 
k(x) = \frac{\left(x^2 - 1\right) \left(\alpha + \gamma\right)}
      {\left(\gamma x + \alpha\right)\left((\alpha + \gamma)(x - 1) + (q - 1)x \right)}\,.
\end{align}
From this expression, it is easy to check that 
\beq
B=\frac{q-1}{2} \left.\frac{d}{dx}K(x)\right|_{x=1}\,.
\eeq
In the next section, we prove the integrability of the $K$-matrix \eqref{K:genHecke} through an algebraic
approach.

\section{Algebraic construction of the boundaries}
\label{sec:algebraicCons}

The integrability of the one-species ASEP can be understood in terms of an underlying Hecke algebra
structure.  From representations of the Hecke and boundary Hecke (or cyclotomic) algebras, solutions of the
Yang-Baxter and reflection equations are constructed through a Baxterisation procedure.  This connection has also
been noted for the two-species ASEP with a certain choice of open boundary conditions \cite{Cantini}.  Indeed, some of
the multi-species ASEP boundary matrices given above fall into the boundary Hecke family.  But in order to encompass all
boundary matrices in the classes \eqref{eq:Bempty} or \eqref{eq:Bfull}, we introduce a new algebra and then show how it is
Baxterised to give solutions of the reflection equation.


\subsection{Hecke algebra \label{sec:heckeb}}

Before presenting the algebra to construct the boundary, let us recall the construction for the R-matrix based on 
the Baxterisation of the Hecke algebra \cite{Jones}.

For $ i=1,2,...,L-1$, we define the following operators
\begin{align}
& \check{\cR}_i(x) = x e_i-e_i^{-1}\;.
 \label{R:heck}
\end{align}
It is well-known that if the generators $e_i$, $1 \le i \le L - 1$ obey the so-called Hecke relations, for $i=1,2,\dots,L-1$ and $j=1,2,\dots,L-2$
\begin{equation}\label{eq:TL2}
    e_i^2 = \omega e_i +1\ , \qquad e_j \,e_{j+1}\, e_j  = e_{j+1}\, e_j\, e_{j+1},
\end{equation}
then the $\check{\cR}(x)$ matrix \eqref{R:heck} obeys the braided Yang-Baxter equation,
\bea
&& \check{\cR}_i(x_1) \check{\cR}_{i+1}(x_1 x_2) \check{\cR}_i(x_2)=
    \check{\cR}_{i+1}(x_2) \check{\cR}_i(x_1 x_2) \check{\cR}_{i+1}(x_1),\label{eq:YBEb}
\eea
and is unitary, up to normalisation.
Using relation $\eqref{eq:TL2}$, the braided R-matrix can be written as follows
\begin{equation}
 \check{\cR}_i(x) = (x-1) e_i+\omega\;.
 \label{eq:heckb}
\end{equation}

One can show that the local Markov matrices $\fm$ provides a representation of the Hecke algebra:
\begin{equation}\label{eq:eiNASEP}
    e_i = (\fm_{i,i+1}+q)/\sqrt{q},
\end{equation}
with $\omega=\sqrt{q}-1/\sqrt{q}$.  Then the $R$-matrix \eqref{eq:Rasep} is written in terms of \eqref{R:heck}
as
\begin{equation}
    R_i(x) = \frac{1}{x\sqrt{q}  - 1/\sqrt{q}} P_{i,i+1} \check{\cR}_i(x).
\end{equation}
The extra factor is necessary for unitarity.  Then relation \eqref{eq:YBEb} implies relation \eqref{ybe}.\\

To summarize, the idea of the Baxterisation \eqref{R:heck} is to get a solution of the Yang-Baxter equation 
(\textit{i.e.} an $R$-matrix depending on a spectral parameter) from a representation of the Hecke algebra.
This idea has been intensively used and generalized to try to classify the solutions of the Yang-Baxter equation
\cite{CGX,ZGB,YQL,BM2000,ACDM,CFRV}.
Then, it has been extended to the reflection equation \cite{LM} through the boundary Hecke algebra \cite{MS}.
However, these algebras are not sufficient to include all the boundary matrices we have constructed.
Below, we  present a slightly more general algebraic structure that encompasses 
all the boundary matrices we found in the previous section, ensuring integrability of the corresponding models.


\subsection{Baxterisation of the K-matrix}

We give in the following proposition the Baxterisation of the K-matrix associated to a Baxterised R-matrix with Hecke algebra.
\begin{proposition}\label{prop:bax}
 Let $e_i$ ($i=1,\dots,L-1$) be the generators of the Hecke algebra satisfying \eqref{eq:TL2} and $\check \cR_i(x)$ the associated braided R-matrices \eqref{R:heck}.
 Let us also define
 \begin{equation}\label{eq:cK}
  \check K(x) = (1-(x-1)e_0)\left(1-\left(\frac{1}{x}-1\right)e_0\right)^{-1}
 \end{equation}
with $e_0$ a supplementary generator.
The inverse in \eqref{eq:cK} is understood as the formal series
\begin{equation}\label{eq:expK}
 \left(1-\left(\frac{1}{x}-1\right)e_0\right)^{-1}=x\,\Big(1-(1-x)(e_0+1)\Big)^{-1}
 =(y+1)\sum_{n=0}^\infty (-y)^n (e_0+1)^n\;,
\end{equation}
where $y=x-1$.

Then $\check K(x)$ is a solution of the braided reflection equation
\begin{equation}\label{eq:bRE}
 \check \cR_1(x_1/x_2) \check K(x_1)\check \cR_1(x_1x_2) \check K(x_2)\ = \ \check K(x_2)\check \cR_1(x_1x_2) \check K(x_1)\check \cR_1(x_1/x_2)
\end{equation}
if and only if the supplementary generator $e_0$ satisfies
 \begin{equation}\label{eq:be}
  e_1\ e_0\ e_1\ e_0\ -\  e_0\ e_1\ e_0\ e_1\ =\ \omega(\ e_0^2\ e_1\ e_0\ -\ e_0\ e_1\ e_0^2\ )\;.
 \end{equation}
 Moreover the $\check K(x)$ matrix is unitary:
\beq
\check K(x)\check K(1/x)=1.
\eeq
\end{proposition}
\proof

We multiply both sides of the braided reflection equation \eqref{eq:bRE} on the left and on the right by 
\begin{equation}
\frac{x_2}{x_1}\left(1-\left(\frac{1}{x_2}-1\right)e_0\right)=\frac1{x_1}(1+(x_2-1)(e_0+1))
\end{equation}
and use \eqref{eq:heckb}, \eqref{eq:cK} 
to get the following equivalent relation 
\begin{eqnarray}
&&(1+y_2(e_0+1))\ ((x_1-x_2)e_1+\omega x_2)\ \frac{1}{x_1}\check K(x_1)\ ((x_1 x_2-1)e_1+\omega)\ (1-y_2 e_0)\nonumber\\
&=&\!\!\!(1-y_2e_0)\ ((x_1x_2-1)e_1+\omega)\ \frac{1}{x_1}\check K(x_1)\ ((x_1-x_2)e_1+\omega x_2)\ (1+y_2(e_0+1))\qquad\label{eq:REeq}
\end{eqnarray}
where $y_i=x_i-1$. Then, we use the expansion \eqref{eq:expK} of $\frac{1}{x_1}\check K(x_1)$ in terms of $y_1$. The coefficient
of $y_1y_2^3$ in \eqref{eq:REeq} provides relation \eqref{eq:be}, which proves that \eqref{eq:bRE} implies \eqref{eq:be}.\\

To prove the reverse implication, we use the following lemma:
\begin{lemma}\label{lem1}
 Relation \eqref{eq:be} implies, for $k=0,1,2,\dots$,
 \begin{eqnarray}
  e_1\ e_0\ e_1\ e_0^k\ -\  e_0^k\ e_1\ e_0\ e_1 &=& \omega(\ e_0^{k+1}\ e_1\ e_0\ -\ e_0\ e_1\ e_0^{k+1}\ ),\label{eq:le1}\qquad\\
   e_1\ e_0^k\ e_1\ e_0\ -\  e_0\ e_1\ e_0^k\ e_1 &=& \omega(\ e_0^{k+1}\ e_1\ e_0\ -\ e_0\ e_1\ e_0^{k+1}\ \label{eq:le2}\\
   &&\hspace{1cm}+\ e_0^{k}\ e_1\ e_0\ -\ e_0\ e_1\ e_0^{k}\ ),\nonumber\\
   e_1\ (e_0+1)^k\ e_1\ e_0 -\ e_0\ e_1\ (e_0+1)^k\ e_1 &=& \omega(\ (e_0+1)^{k+1}\ e_1\ e_0\ \nonumber\\
   &&\hspace{1cm}-\ e_0\ e_1\ (e_0+1)^{k+1}\ )\;,\label{eq:le3}\\
   e_1\ e_0\,(e_0+1)^k\ e_1\ e_0 -\ e_0\ e_1\ e_0\,(e_0+1)^k\ e_1 &=& \omega(\ e_0\ (e_0+1)^{k+1}\ e_1\ e_0\ \nonumber\\
   &&\hspace{1cm}-\ e_0\ e_1\ e_0\ (e_0+1)^{k+1}\ )\;.\ \label{eq:le4}
 \end{eqnarray}
\end{lemma}
The first relation of the lemma \eqref{eq:le1} is proven by recursion using \eqref{eq:be}. Relation \eqref{eq:le2} is proven also by
recursion with \eqref{eq:le1} and \eqref{eq:TL2}. The third and the fourth are proven  by expanding $(e_0+1)^k$ and using \eqref{eq:le2}.

The lemma allows us to prove that
\begin{equation}\label{eq:ek}
 e_1\check K(x)e_1e_0-e_0e_1\check K(x_1)e_1=\omega\left((e_0+1)\check K(x)e_1e_0-e_0e_1(e_0+1) \check K(x)   \right)\;.
\end{equation}

Finally, by expanding \eqref{eq:REeq} and by using relation \eqref{eq:ek}, we prove that equation \eqref{eq:be}
implies \eqref{eq:bRE} which concludes the proof of the proposition.
\finproof

\paragraph{Connection with Baxterisation of cyclotomic Hecke algebras.} Another Baxterisation for the K-matrix was proposed in \cite{KM}, starting 
from a slightly different algebra. There, the relation \eqref{eq:be} is replaced by
\begin{eqnarray}\label{eq:KM}
&&  e_1\, \bar e_0\, e_1\, \bar e_0\ -\  \bar e_0\, e_1\, \bar e_0\, e_1\ =\ 0\\
 && \sum_{k=0}^m a_k\, (\bar e_0)^k=0\label{cyclo}
 \end{eqnarray}
for some fixed $m=2,3,\dots$ and $a_0$, ... $a_m$ free parameters. 
The relation \eqref{cyclo} is called the cyclotomic relation.
Then, a K-matrix can be constructed as a polynomial in $\bar e_0$ \cite{KM}.
When $m=2$, the cyclotomic Hecke algebra is just the boundary Hecke algebra.

In fact, similarly to proposition \ref{prop:bax}, one can show that
\begin{equation}\label{eq:Kcyclo}
 \check K(x)=(1-x\bar e_0)\left(1-\frac1x\bar e_0\right)^{-1}
\end{equation}
 satisfies the reflection equation, provided $\bar e_0$ satisfies solely the relation \eqref{eq:KM}. 
The polynomial Baxterisation of \cite{KM} is recovered when one assumes in addition the cyclotomic relation \eqref{cyclo}.

One can match this Baxterisation  with the one
 presented in \eqref{eq:cK} in the following way. Starting from the algebra \eqref{eq:be}, and assuming that $(e_0+1)$ is invertible, it is possible to prove that the generator
\begin{equation}\label{eq:moKM}
 \bar e_0=e_0 (1+e_0)^{-1}
\end{equation}
satisfies the relation \eqref{eq:KM}. This can be shown by using relation \eqref{eq:be} for $e_0$ and lemma \ref{lem1}. 
Then, substituting \eqref{eq:moKM} into the  Baxterised K-matrix \eqref{eq:Kcyclo} yields \eqref{eq:cK} up to a normalisation factor. 

\subsection{Integrability of the multi-species ASEP boundary matrices\label{sec:nasepInteg}} 
The aim of this section is to prove that the boundary matrices presented in section \ref{sect:solu} fit into the Baxterisation procedure of proposition \ref{prop:bax}.

\begin{proposition}
For any matrix $B = B(\alpha, \beta | s_1, s_2, f_2, f_1)$ or $B = B^0(\alpha, \beta | s_1, s_2, f_2, f_1)$, the generators
\begin{equation}\label{eq:e0NASEP}
    e_0 = \frac{B + \alpha + \gamma + q - 1}{1 - q}
    \mb{and} e_1 = (\fm+q)/\sqrt{q}
\end{equation}
obey relation \eqref{eq:be}, where $\fm\equiv \fm_{12}$ is given in \eqref{eq:mlocal} and $B$ acts non trivially in space 1.
\end{proposition}
\proof
The matrices $e_1$ and $e_0$ given in \eqref{eq:eiNASEP} and \eqref{eq:e0NASEP} act
on two site multi-species ASEP configurations.
For a given start state, $\tau_1 \tau_2$, we can find a subset of the particle species
$\cS = \{\tau_1, \tau_2, \tau_3, \ldots \}$ such that for \textit{any} polynomial in $e_1$ and $e_0$
acting on this state, these are the only species involved in the resulting configurations.

For all of the boundary matrices we consider, the subset $\cS$ turns out to be small, and related to the
different classes of particles we introduced above: the non-diagonal part of $e_1$ exchanges particles on
sites 1 and 2, as allowed by bulk matrix $\fm$; the non-diagonal part of $e_0$ injects and removes particles
at site 1 as allowed by the boundary transitions given in section \ref{sect:solu}.  The idea of the proof is
then to project the `global' matrices $e_0$, $e_1$ down to the smaller number of species in $\cS$.  If for
every starting state we can show that the resulting projected $e_0$, $e_1$ satisfy \eqref{eq:be}, then
this implies that the `global' matrices also satisfy \eqref{eq:be}.

At this point, the proof decomposes into different steps:
\begin{itemize}
    \item
We remark that for any start state $\tau_1 \tau_2$, the set $\cS$ falls into one of three categories:
\bea
&& \mathcal{S} = \{\tau_1, \tau_2, s_1,s_2, f_1,f_2\},\\
&& \mathcal{S} = \{\tau_1, s_1+ f_1-\tau_1,\tau_2,s_1+ f_1-\tau_2\},\\
&& \mathcal{S} = \{\tau_1, s_1+ f_1-\tau_1,\tau_2, s,f\},\mb{with} (s,f)=(s_1, f_1) \mbox{ or } (s_2, f_2)
   \label{eq:Smixed}
\eea
Note that these sets can be reduced depending on the class of the species $\tau_1$ and $\tau_2$. For instance,
if $\tau_1$ and $\tau_2$ are of very slow class, then $\mathcal{S} = \{\tau_1, \tau_2, s_1,f_1\}$.  Note also
that the ordering of the start state does not change $\cS$ so $\tau_1$, $\tau_2$ are interchangeable in
\eqref{eq:Smixed}.
    \item
Projecting the boundary matrix, $B$, corresponding to $e_0$ down to the species in $\cS$ results in a boundary
matrix of size $|\cS|$ of type \eqref{eq:Bempty} or \eqref{eq:Bfull}.  To see this, we perform the projection
by `deleting' species from $B$ by removing the corresponding row and column: 
 we  use  the following operations which preserve the forms \eqref{eq:Bempty} or \eqref{eq:Bfull}:
    \begin{itemize}
        \item
    Deleting any species in the very slow, intermediate, or very fast class;
        \item
    Deleting a species, $\tau$, in the slow or fast class with $\tau \ne s_1, f_1$ if we also delete
    the species $s_1 + f_1 - \tau$;
        \item
    Deleting species $s_1$ and $f_1$ together, if $s_1 = 1$, $f_1 = N$, and $f_1 - f_2 = s_2 - s_1 > 0$.
        \item
    Deleting species $s_2$ and $f_2$ together, if $f_2 = s_2 + 1$, and $f_1 - f_2 = s_2 - s_1 > 0$.
    \end{itemize}
    These operations are always sufficient to project down to any subsets $\cS$ as defined above.
The projected $e_0$ is then obtained from the projected $B$ through \eqref{eq:e0NASEP}.
    \item
For the local bulk matrix $\fm$ (giving $e_1$) we can delete any number of species, preserving the form
\eqref{eq:mlocal}.
    \item
To complete the proof all we need to do is to verify that all boundary matrices in this family give $e_0$
matrices which satisfy  \eqref{eq:be} for size $2$ up to $6$ (the maximum $|\cS|$).  We have done this by a direct
computation with a formal mathematical software package.
\end{itemize}
To illustrate the projection on $\cS$, we consider the following boundary matrix
\bea
&&B=
\left(
\begin{array}{c|cc|c|cc|c}
 \mbox{-}\sigma & & & & & &
\\
\hline
\gamma & \mbox{-}\alpha && & &\wt\gamma & \wt\gamma
\\
 & &\mbox{-}\alpha & \wt\gamma & \wt\gamma & &
\\
\hline
 & & & \mbox{-}\sigma' & &&
\\
\hline
 & &\alpha & \alpha & \mbox{-}\wt\gamma & &
\\
 \alpha& \alpha& & & &\mbox{-}\wt\gamma & \wt\alpha
 \\
 \hline
 & & & & & & -\wt\sigma
\end{array}
\right)
\eea
and give some examples of start state $(\tau_1,\tau_2)$ and the resulting subset $\cS$ and corresponding reduced matrix.
In the case where $(\tau_1,\tau_2)=(1,4)$, we obtain $\cS=\{1,4,s_1=2,s_2=3,f_1=6,f_2=5\}$ and the reduced matrix reads
\beq
\left(
\begin{array}{c|cc|c|cc}
 \mbox{-}\sigma & & & & & 
\\
\hline
\gamma & \mbox{-}\alpha && & &\wt\gamma 
\\
 & &\mbox{-}\alpha & \wt\gamma & \wt\gamma & 
\\
\hline
 & & & \mbox{-}\sigma' & &
\\
\hline
 & &\alpha & \alpha & \mbox{-}\wt\gamma & 
\\
 \alpha& \alpha& & & &\mbox{-}\wt\gamma 
\end{array}
\right).
\eeq
In the case where  $(\tau_1,\tau_2)=(2,6)$, we obtain $\cS=\{2,6\}$ and the reduced matrix reads
\beq
\begin{pmatrix} \mbox{-}\alpha & \wt\gamma \\ \alpha & \mbox{-}\wt\gamma \end{pmatrix}.
\eeq
Finally, in the case where $(\tau_1,\tau_2)=(3,4)$, we obtain $\cS=\{3,4,5\}$ and the reduced matrix reads
\beq
\begin{pmatrix} \mbox{-}\alpha & \wt\gamma& \wt\gamma \\ 
0 &\mbox{-}\sigma' &0 \\ \alpha & \alpha & \mbox{-}\wt\gamma \end{pmatrix}.
\eeq
\endproof

From proposition \ref{prop:bax}, the generator $e_0$ defined above provides a $\check K(x)$ matrix that obeys the reflection  equation and is unitary. The connection with the expression \eqref{K:genHecke} is given by the following proposition. 
\begin{proposition}
For any matrix $B = B(\alpha, \beta | s_1, s_2, f_2, f_1)$ or $B = B^0(\alpha, \beta | s_1, s_2, f_2, f_1)$, let $B= b_0 + b_0^+ + b_0^-$ be the decomposition described in section \ref{sect:Bdecomp}.
The matrix 
\begin{align}
    & K(x) = 1 + k(x)\,\Big(b_0 + x \,b_0^+ + \frac1x \,b_0^-\Big),\label{K:genHecke2}\\
    & \mbox{with}\quad 
k(x) = \frac{\left(x^2 - 1\right) \left(\alpha + \gamma\right)}
      {\left(\gamma x + \alpha\right)\left((\alpha + \gamma)(x - 1) + (q - 1)x \right)}\,
\end{align}
can be expressed as  a Baxterised $\check K(x)$ matrix
\begin{equation}\label{eq:Knasep}
    K(x) = \frac{(\alpha + \gamma + q - 1)(\frac{1}{x} - 1) + q - 1}{(\alpha + \gamma + q - 1)(x - 1) + q - 1}
           \left(\frac{1 - (x - 1)e_0}{1 - (\frac{1}{x} - 1) e_0}\right),
\end{equation}
where
\begin{equation}
    e_0 = \frac{B + \alpha + \gamma + q - 1}{1 - q}.
\end{equation}
Thus, it satisfies the reflection equation and is unitary.
\end{proposition}
\proof
Note that $K(x)$ in \eqref{eq:Knasep} has the form
\begin{equation*}
    K(x) = \frac{f(x)}{f(1/x)} \check{K}(x)
\end{equation*}
with $\check{K}(x)$ as in \eqref{eq:cK} so that it remains unitary and satisfies the reflection equation.  

To show that the $K$-matrix \eqref{eq:Knasep} is equivalent to the form \eqref{K:genHecke2} we will need the
following relations for the matrices $b_0$, $b_0^+$ and $b_0^-$:
\begin{equation}\label{poly:e0b}
\begin{aligned}
    & b_0^2 = -\left(\alpha + \tilde{\gamma}\right) b_0 + \tilde{\alpha} b_0^+ + \gamma b_0^-,
      \qquad
      \left(b_0^+\right)^2 = - \gamma b_0^+,
      \qquad
      \left(b_0^-\right)^2 = -\tilde{\alpha} b_0^-,
      \\
    & b_0 b_0^+ = b_0^+ b_0 = -\alpha b_0^+,
      \qquad
      b_0 b_0^- = b_0^- b_0 = -\tilde{\gamma} b_0^-,
      \\
    &
      b_0^+ b_0^- = b_0^- b_0^+ = 0.
\end{aligned}
\end{equation}
They involve the combination of parameters defined in \eqref{eq:tilde}. 
These relations are proven
 using the same method of projecting down to a set $\cS$ of all species involved from a given starting
state.  In this case, the matrices act on a single site configuration, and it is sufficient to check that the
relations hold for size $2$ and $3$.

Using \eqref{poly:e0b} it is straightforward to check that
\begin{eqnarray}
&& \left(1-\left(\frac{1}{x}-1\right)e_0\right)\,
\left(q-1+\frac{(x-1)(\alpha+\gamma)}{(\alpha+\gamma x)}\left(\frac{b_0^{-}}{x}+b_0+b_0^{+}x\right)\right)
\\
&&=\frac1{x}(\alpha+\gamma+q-1-(\alpha+\gamma)x),
\end{eqnarray}
which allows $\left(1-\left(\frac{1}{x}-1\right)e_0\right)^{-1}$ to be expressed as a polynomial in $b_0,b_0^\pm$.
Then using \eqref{poly:e0b} again we can show that
\begin{eqnarray}
&& \frac{(\alpha + \gamma + q - 1)(\frac{1}{x} - 1) + q - 1}{(\alpha + \gamma + q - 1)(x - 1) + q - 1}
           \Big(1 - (x - 1)e_0\Big)\left(1 - (\frac{1}{x} - 1) e_0\right)^{-1}= \\
&& 1 + \frac{\left(x^2 - 1\right) \left(\alpha + \gamma\right)}
      {\left(\gamma x + \alpha\right)\left((\alpha + \gamma)(x - 1) + (q - 1)x \right)}\,\Big(b_0 + x \,b_0^+ + \frac1x \,b_0^-\Big),
\end{eqnarray}
which concludes the proof.

\endproof

\paragraph{Polynomial relations.} 
Using relations \eqref{poly:e0b} and the expression \eqref{eq:e0NASEP} for $e_0$, we get 
\begin{equation}\label{eq:anne}
 e_0(e_0+1)\left(e_0+\frac{\alpha}{\alpha+\gamma}\right)\left(e_0+\frac{\alpha+\gamma+q-1}{q-1}\right)=0\;.
\end{equation}
%
We stress  that the factor $(e_0+1)$ is present in \eqref{eq:anne}. 
Moreover, we find that for particular choices of $b_0$, $b_0^+$ and $b_0^-$, the polynomial 
\eqref{eq:anne} becomes minimal for $e_0$.
Then, in these cases, $(e_0+1)$ is not invertible and we cannot use 
\eqref{eq:moKM} to Baxterise the K-matrix for the multi-species ASEP from the construction of \cite{KM}.
However, the model is still integrable thanks to the Baxterisation \eqref{eq:cK}.

Note that when $b_0^+=b_0^-=0$, relations \eqref{poly:e0b}  reduce to $b_0(b_0+\alpha+\wt\gamma)=0$ and we recover the boundary Hecke algebra.

\section{Conclusion and perspectives}

In this paper, we present integrable boundary matrices for the multi-species asymmetric exclusion process.
We believe that the solutions presented here are the only Markovian solutions of the reflection equation 
with at least two free parameters\footnote{We have found other distinct solutions, but they have only one free
parameter and are physically less interesting.}.
This conjecture is supported by two facts:
$(i)$ we recover the classification done for the one and the two-species models;
$(ii)$ we also checked that for the three-species case, and
assuming that the K-matrix entries are  polynomials of degree 4 w.r.t. the spectral parameter,
the only solutions to the reflection equation are the ones presented in equation \eqref{K:genHecke} which are of degree 3 (up to a normalisation).

As explained previously, these solutions allow us to define integrable Markovian stochastic processes. Then, we believe 
that the associated stationary state can be expressed with a matrix ansatz following the generic idea developed in \cite{SW,CRV} 
and already exploited for the two-species case in \cite{CMRV}. The integrability of these models should permit
also the computation of other physical 
quantities such as correlation functions, the spectral gap, and fluctuations of the density and of the
current.

The second result of this paper provides an algebraic framework for these solutions which generalize the boundary 
Hecke algebra. We hope that this algebra can be exploited in other contexts such as quantum integrable spin
chains or the $O(1)$ loop model, allowing one to generalize the results described in \cite{doikou,MNGB}.
The boundary Hecke algebra has also been used to relate some stationary weights of open semi-permeable two-species ASEP to Koornwinder polynomials \cite{Cantini}.
The algebraic structure presented here may be relevant to extend this study to the case of models with permeable integrable boundaries.

\section*{Acknowledgments}
We are grateful to Jan de Gier for helpful discussions.


\begin{thebibliography}{99}

\bibitem{abad} J. Abad and M. Rios, 
\textsl{Nondiagonal solutions to reflection equations in SU(N) spin chains}, 
Phys. Lett. \textbf{B 352} (1995) 92 and \texttt{hep-th/9502129}.

\bibitem{arita} C. Arita,
\textsl{Remarks on the multi-species exclusion process with reflective boundaries,}
J. Phys. \textbf{A 45} 155001 (2012) and \texttt{arXiv:1112.5585}.

\bibitem{AACDFR} D. Arnaudon, J. Avan, N. Crampe, A. Doikou, L. Frappat and E. Ragoucy,
\textsl{Classification of reflection matrices related to (super) Yangians and application to open spin chain models,}
Nucl. Phys. \textbf{B 668} (2003) 469 and \texttt{math.QA/0304150};\\
id. \textsl{General boundary conditions for the $sl(N)$ and $sl(M|N)$ open spin chains,}
J. Stat. Mech. (2004) P08005 and \texttt{math-ph/0406021}.

\bibitem{ACDM} D. Arnaudon, A. Chakrabarti, V.K. Dobrev and  S.G. Mihov, 
\textsl{Spectral Decomposition and Baxterisation of Exotic Bialgebras and Associated Noncommutative Geometries,}
Int. J. Mod. Phys. \textbf{A 18} (2003) 4201 and \texttt{arXiv:math/0209321}.

\bibitem{BFKZ} M.T. Batchelor, V. Fridkin, A. Kuniba and Y.K. Zhou, 
\textsl{Solutions of the reflection equation for
face and vertex models associated with $A_n^{(1)}$, $B_n^{(1)}$, $C_n^{(1)}$, $D_n^{(1)}$ and $A_n^{(2)}$},
Phys. Lett. \textbf{B 376} (1996) 266 and \texttt{hep-th/9601051}.

\bibitem{BM2000} S. Boukraa and J.M. Maillard,
\textsl{Let's Baxterise,}
J. Stat. Phys. \textbf{102} (2001) 641 and \texttt{arXiv:hep-th/0003212}.

\bibitem{Cantini} L. Cantini,
\textsl{Asymmetric Simple Exclusion Process with open boundaries and Koornwinder polynomials},
\texttt{arXiv:1506.00284}.

\bibitem{CGW} L. Cantini, J. de Gier and M. Wheeler,
\textsl{Matrix product and sum rule for Macdonald polynomials},
\texttt{arXiv:1602.04392}.

\bibitem{CGX} Y. Cheng, M.L. Ge and K. Xue, 
\textsl{Yang--Baxterization of Braid Group Representations,} 
Commun. Math. Phys. \textbf{136} (1991) 195.

\bibitem{CMZ} T. Chou, K. Mallick and R. K. P. Zia,
\textsl{Non-equilibrium statistical mechanics: From a paradigmatic model to biological transport,}
Rep. Prog. Phys. \textbf{74} (2011) 116601 and \texttt{arXiv:1110.1783}.

\bibitem{CSS} D. Chowdhury, L. Santen and A. Schadschneider,
\textsl{Statistical Physics of Vehicular Traffic and Some Related Systems},
Phys. Rep. \textbf{329} (2000) 199 and \texttt{arXiv:cond-mat/0007053}.

\bibitem{Corteel} S. Corteel, O. Mandelshtam and L. Williams,
\textsl{Combinatorics of the two-species ASEP and Koornwinder moments},
\texttt{arXiv:1510.05023}.

\bibitem{CFRV}N. Crampe, L. Frappat, E. Ragoucy and M. Vanicat,
\textsl{A new braid-like algebra for Baxterisation,}
\texttt{arXiv:1509.05516}.

\bibitem{CMRV} N. Crampe, K. Mallick, E. Ragoucy and M. Vanicat,
\textsl{Open two-species exclusion processes with integrable boundaries,}
J. Phys. \textbf{A 48} (2015) 175002 and \texttt{arXiv:1412.5939}. 

\bibitem{CRV} N. Crampe, E. Ragoucy and M. Vanicat,
\textsl{Integrable approach to simple exclusion processes with boundaries. Review and progress,}
J. Stat. Mech. (2014) P11032 and \texttt{arXiv:1408.5357}.

\bibitem{gier}J. de Gier and F.H.L. Essler,
\textsl{Exact Spectral Gaps of the Asymmetric Exclusion Process with Open Boundaries,}
 J. Stat. Mech. (2006) P12011 and \texttt{arXiv:cond-mat/0609645}.

\bibitem{Derrida98} B. Derrida
\textsl{An exactly soluble non-equilibrium system: The asymmetric simple exclusion process,}
Phys. Rep. \textbf{301} (1998) 65.

\bibitem{vega} H.J. de Vega and A. Gonz\'{a}lez-Ruiz, 
\textsl{Boundary K matrices for the six vertex and the $n(2n-1)$ $A_{n-1}$ vertex models}, 
J. Phys. \textbf{A 26} (1993) L519 and \texttt{hep-th/9211114}.

\bibitem{doikou} A. Doikou,
\textsl{Boundary non-local charges from the open spin chain,}
J. Stat. Mech. (2005) P12005 and \texttt{arXiv:math-ph/0402067}.

\bibitem{gan} G.M. Gandenberger, 
\textsl{New non-diagonal solutions to the $a_n^{(1)}$ boundary Yang-Baxter equation,}
\texttt{hep-th/9911178}.

\bibitem{cumulants} M. Gorissen, A. Lazarescu and K. Mallick, 
\textsl{Exact Current Statistics of the ASEP with Open Boundaries,} 
Phys. Rev. Lett. \textbf{109} (2012) 170601 and \texttt{arXiv:1207.6879}.

\bibitem{Jones} V.F.R. Jones, 
\textsl{Baxterisation,}
Int. J. Mod. Phys. \textbf{B 4} (1990) 701, proceedings of ``Yang-Baxter equations, 
conformal invariance and integrability in statistical mechanics and field
theory'', Canberra, 1989.

\bibitem{KM} P.P. Kulish and  A.I. Mudrov,
\textsl{Baxterization of solutions to reflection equation with Hecke R-matrix,}
Lett. Math. Phys. \textbf{75} (2006) 151 and \texttt{arXiv:math/0508289}.
    
\bibitem{KMO} A. Kuniba, S. Maruyama and M. Okado, 
\textsl{Multispecies TASEP and the tetrahedron equation},
J. Phys. \textbf{A 49} (2016) 114001 and \texttt{arXiv:1509.09018}.    

\bibitem{LM} D. Levy and P. Martin,
\textsl{Hecke algebra solutions to the reflection equation,}
J. Phys. \textbf{A 27} (1994) L521.

\bibitem{YQL} Y.-Q. Li,
\textsl{Yang Baxterization,}
J. Math. Phys. \textbf{34} (1993) 757.

\bibitem{Liggett} T.M. Liggett,
\textsl{Interacting Particle Systems,}
Springer, New York, 1985.

\bibitem{PhD-Corteel} O. Mandelshtam,
\textsl{Matrix ansatz and combinatorics of the $k$-species PASEP,}
\texttt{arXiv:1508.04115}.

\bibitem{MS} P.P. Martin and H. Saleur, 
\textsl{On an algebraic approach to higher-dimensional statistical mechanics}, 
Comm. Math. Phys. \textbf{158} (1993) 155 and \texttt{hep-th/9208061};\\
id. \textsl{The blob algebra and the periodic Temperley-Lieb algebra,} 
Lett. Math. Phys. \textbf{30} (1994) 189 and \texttt{hep-th/9302094}.


\bibitem{MNGB} S. Mitra, B. Nienhuis, J. de Gier and M.T. Batchelor, 
\textsl{Exact expressions for correlations in the ground state of the dense O(1) loop model,} 
J. Stat. Mech. (2004) P09010 and \texttt{cond-mat/0401245}.

 \bibitem{pem} S. Prolhac, M. R. Evans and K. Mallick,
 \textsl{Matrix product solution of the multispecies partially asymmetric exclusion process,}
 J. Phys. \textbf{A 42} (2009) 165004 and \texttt{arXiv:0812.3293}.
    
\bibitem{SW} T. Sasamoto and M. Wadati,
\textsl{Stationary state of integrable systems in matrix product form,}
J. Phys. Soc. Japan \textbf{66} (1997) 2618.    

\bibitem{Spitzer} F. Spitzer
\textsl{Interaction of Markov processes,}
Advances in Mathematics \textbf{5} (1970) 246.

\bibitem{skl} E.K. Sklyanin, 
\textsl{Boundary conditions for integrable quantum systems,} 
J. Phys. \textbf{A 21} (1988) 2375.
  
\bibitem{Uchiyama}  M. Uchiyama,
\textit{Two-Species Asymmetric Simple Exclusion Process with Open Boundaries,}
Chaos, Solitons \& Fractals \textbf{35}, (2008) 398 and \texttt{arXiv:cond-mat/0703660}.

\bibitem{ZGB} R.B. Zhang, M.D. Gould and A.J. Bracken, 
\textsl{From representations of the braid group to solutions of the Yang--Baxter equation,}
Nucl. Phys.  \textbf{B 354} (1991) 625.



\end{thebibliography}
\end{document}